# Let us play with qubits

Sylvain Gravier*, Philippe Jorrand* , Mehdi Mhalla†and Charles Payan*

5th November 2003

## 1 Introduction

Quantum game theory offers a lot of interesting questions, and it is relevant to use the quantum information theory to resolve or improve games with lack of information : how to use the power of quantum entanglement to show the superiority of a quantum player that is allowed to use quantum mechanics versus a classical player (penny flipover [5]...), how to use quantum communication properties in cooperative games (prisoners game, [3], guessing number [8],...). An introduction to this field and an overview of important results is proposed in [6, 7]. But games are also useful to make notions easier to understand, and permit to apprehend easier new ways of reasoning. The objective of this work is to formalize and to study a simple game with qubits using quantum notions of measurement and superposition but keeping a simple formalism so that knowing quantum mechanics is not necessary to play the game. We solve a quantum combinatorial game $Q.007$ by giving a winning strategy for it. We also propose a quantisation of a family of combinatorial games. A playable version of the studied qubit game is available at the address *http://www-leibniz.imag.fr/QUI/demo/testjeu.html*.

## 2 Definition

We have a graph with vertices that can take four colors : white, light gray, black and dark gray. Initially, all vertices are white. The game consists in choosing a vertex, if it is a black one, the player looses and the game stops, if it is a grey one, the player has a 1/2 probability to loose, then the player removes it from the graph and change the color of the neighbor's following the scheme : White → Light Gray → Black → Dark Gray → White. If there is no more vertices the player looses.

*CNRS, Laboratoire Leibniz, 46 avenue Félix Viallet, 38031 Grenoble Cedex, France
†CNRS, Doctoral Fellow, Laboratoire Leibniz, 46 avenue Félix Viallet, 38031 Grenoble Cedex, France

This definition may seem unnatural but it has a natural quantum definition: each vertex represents a qubit, white color corresponds to the state $|0>$, light gray corresponds to the state $\sqrt{Not}(|0>)$, black corresponds to $|1>$ and dark gray to $\sqrt{Not}(|1>)$. So the game consists in choosing a qubit measuring it, removing it from the game and applying the gate $\sqrt{Not}$ to its neighbors. The first player that measure the state $|1>$ or that can not play looses.

This game can also be seen as a two players game version of the Sutner's sigma game (recently generalized in [4]).

If we consider the game over a chain, we can represent a vertex. Two squares are adjacent iff they meet along an edge.

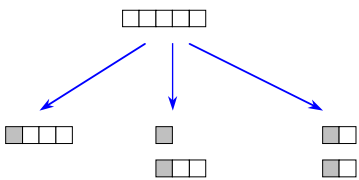

Figure 1: All possible plays on a chain of 5 vertices

## 3 Auxiliary Games

In this section, we study some classical games, that will permit to analise the quantum version. The main properties (sum game, equivalence classes) are due to Grundy and Spragues, our contribution consists in adding a precise description of the equivalence classes of the game and exhibiting an optimal strategy.

### 3.1 White Game

The first phase of analysis consists in considering only the no risk vertices (white ones). So we can say that this game (white game) consists in removing a vertex and its neighbors, the first player that ca not play looses.

### 3.2 Domino Game

The classical domino game's rule is: each player removes a domino (two adjacent vertices). The first player who cannot play looses.

### 3.3 Relationship between Domino Game and White Game

**Lemma 1 :**

*The domino game over a row of lenght $n$ is equivalent to the white game over a row of lenght $n + 1$.*

**proof :**

If the considered graph is a chain of $n$ vertices ($n-1$ edges), playing the domino game on the vertices (removing two vertices) is equivalent to playing the white game on the edges (removing an edge and the incident edges). □

## 3.4 Study of the Domino Game on a Chain

In the general case, determining if there exists a winning strategy is an open problem (even for grids), we will focus our attention in the chain case.

### 3.4.1 Playing Situations

A playing situation is a set of chains, it is an intermediate state of the game, it is obtained by playing over an initial chain. A player's turn is a transition from a game situation to a possible next situation. We can define the digraph (oriented graph) of transitions $TG$. As the number of playable squares strictly decreases, the transitions induce a partial order over the situations so $TG$ has no circuit. Some vertices of $TG$ corresponds to winning situations: there exists a winning strategy starting from such situations, the other vertices corresponds to loosing situations. A terminal loosing situation is one that has no transitions out of it (in the game: a set, which can be empty, of isolated vertices): indeed a player in such a situation can't play, they are minimal situations for the partial order induced by the transitions. All situations that have at least one transition to a loosing situation are winning ones (we can put the other player in a loosing situation). We note $W$ the set of winning situations and $L$ the set of loosing ones.

### 3.4.2 Sum Game

A Sum game is a game on a couple of situations: a player can play on any of both situations and he looses if he cannot play in any of them so terminal loosing situation for the sum game is the sum of two terminal loosing situations.

For the domino game, a pair of situations is a set of rows and so it is also a game situation. We have the natural properties:

$$\begin{aligned} if\ S_1 \in W, \quad &S_2 \in L,\ S_3 \in L \\ &S_1 + S_2 \in W \\ &S_2 + S_3 \in L \end{aligned} \quad (1)$$

The sum game allows us to define a relation between situations: we say that $X$ and $Y$ are in the same equivalence class if $X + Y$ is a loosing situation. It is an equivalence relation:

It's a symmetric relation : $X + Y \in L \Leftrightarrow Y + X \in L$

Reflexitivity :

$X + X$ is always a loosing situation because of "mimic" play: A product of terminal loosing situation is in $L$ and we always can have by transitions $X + X \to X + Y \to Y + Y$ by playing in the second situation the domino that is symmetric to the one chosen in the first situation, so by induction all the $X + X$ are in $L$.

Transitivity : If $X + Y \in L, Y + Z \in L$ then $X + Z \in L$

By way of contradiction suppose $X + Z \in W$, then by 1 $((X + Z) + (Y + Y)) \in W$. But these sets of rows can be seen as the sum situation $((X + Y) + (Y + Z))$ and so we get: $((X + Y) + (Y + Z)) \in W$, which is a contradiction by 1.

So it's an equivalence relation.

### 3.4.3 Class numbering

A kernel of a graph $G = (V, E)$ is a subset $C$ of $V$ such that, $C$ is a **stable** (no edges in $G$ between the vertices of $C$) and is a **dominating set** : each vertex in $V \setminus C$ has an edge from it to $C$. It is well known that a graph without circuits admits a unique kernel [1]. Since the transition graph has no circuits it admits a unique kernel, lets denote it $C_0$. Now we see that $C_0$ corresponds to loosing situations. Indeed, first remark that all terminal situations are in $C_0$, and a winning situation have a transition to $C_0$ (by domination) that corresponds to putting the other player in a loosing situation whereas a loosing situation have no transition to $C_0$ (it is a stable).

By removing $C_0$ from the graph we can find a new kernel $C_1$ in the rest of the graph. We define recursively the classes $C_n$ using this process. A situation in $C_n$ can reach by one transition all the classes $C_k$ with $k < n$ (because of the domination property of $C_k$), so for every situation $X$ in $C_n$, $n$ will be the smallest index of class that $X$ can not reach in one transition.

**Lemma 2** :

If $X \in C_i$ and $Y \in C_j$ then $X + Y \in C_{i \oplus j}$

**proof** :

By 1 and by definition of $C_0$ we have: if $X \in C_0$ and $Y \in C_0$ then $X + Y \in C_0$.

Let $X \in C_i$ and $Y \in C_j$, suppose the property holds for all situations $X' + Y'$ smaller then $X + Y$.

First, we prove that there is a transition from $X + Y$ to some $Z \in C_k$ for every $k < i \oplus j$.

Let $k < (i \oplus j)$, and $i_0$ the index of the first different binary bit between $k$ and $i \oplus j$ so :

$$\begin{aligned} (i \oplus j) &= s1s' \\ k &= s0k' \end{aligned}$$

Without loss of generality, we assume that the $i_0^{th}$ binary bit of $i$ is 1 and the $i_0^{th}$ binary bit of $j$ is 0, we have :

$$\begin{aligned} i &= u1i' \\ j &= v0j' \end{aligned}$$

Where $s = u \oplus v$ and $X$ can reach by one transition a situation $Z$ in the class $u0(k' \oplus j')$ (smaller class) so that the class index of $Z + Y$ will be $(u \oplus v)0(k' \oplus j' \oplus j') = k$.

So the index of the class of $X + Y$ is greater or equal than $(i \oplus j)$.

Moreover, the class $i \oplus j$ is not reachable from $X + Y$ by one transition : We can suppose without loss of generality that we play on the situation $Y$ in $C_j$ that becomes a situation $Z$ in $C_k$. By induction, the class of $X + Z$ is $i \oplus k$. Suppose $i \oplus k = i \oplus j$, we would have $k = j$ and so a transition from $C_j$ to $C_j$ impossible. So $X + Y \in C_{i \oplus j}$.

All these results could be easily deduced from [2] . □

**Corollary 1 :**

The classes $C_i$ are the classes of the equivalence relation of the sum game.

**proof :**

$X$ and $Y$ are in the same class $C_i$ iff $X + Y$ is a loosing situation (in $C_0 = C_{i \oplus i}$). □

### 3.4.4 Strategy for winning cases

Let a situation $S = \{B_1, \ldots, B_n\}$ be in the class $s$, and let $c_i$ be the index class of the $i^{th}$ row: $s = c_1 \oplus c_2 \oplus \ldots \oplus c_n$. If the set is winning situation ($s \neq 0$) then $s$ can be written in binary: $0..01s'$. Let $i_0$ be the index of the first 1 (counting from the left). There exists a row $B_i$ such that $c_i$ , written in binary, has 1 in position $i_0$: $c_i = u1v$. A winning transition is one that brings this row $B_i$ from the class $c_i$ to the class $c'_i = u0(s' \oplus v)$ which corresponds to a reachable class because $c'_i$ is smaller than $c_i$. Therefore, the new class of the set becomes $s \oplus ci \oplus c'_i = 0$.

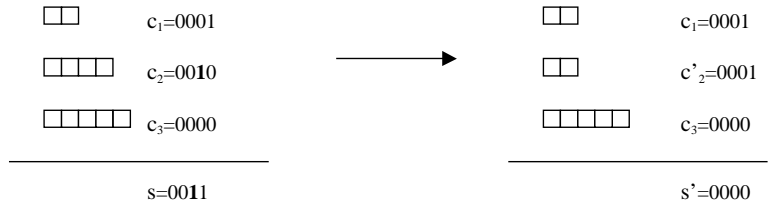

Figure 2: Example of winning transition

### 3.4.5 Row classes

A row of length 0 is loosing ( no playing transitions) so it is in $C_0$.

A row of length 1 is loosing ( no playing transitions) so it is in $C_0$.

A row of length 2 can give by one transition a row of length 0 (in $C_0$) thus it is in $C_1$.

A row of length 3 can give by one transition a row of length 1 (in $C_0$) thus it is in $C_1$.

A row of length 4 can give by one transition a row of length 2 (in $C_1$) or two rows of length 1 (in $C_0$) so it is in $C_2$.

We can repeat this process using the property that a row of length $n$ can give by one transition : a row of length $n-2$ or two rows, one of length $k$ and the other of length $n-2-k$ where $0 < k < n-3$ ( by symmetry we can take $k \leq \lfloor n/2 \rfloor - 1$), thus by lemma 2 we can determine a row's class knowing the classes of smaller ones. This classification of rows depending on their length make appear the pseudo-period 34.

We can notice that except for the rows 0,1,15,35 (that are in $C_0$) and 17,18,32,52 (that are in $C_2$) it is sufficient to determine the modulo 34 of the length of a row to determine its class using the second column.

The classes $C_6, C_{10}, C_{11}, C_{12}, C_{13}, C_{14}, C_{15}$ contain only sets of rows : no single row is in these classes. To prove the correctness of the table 1, we compute the values of the classes for the rows of length smaller than 176, then lemma 3 ensures the correctness for larger rows.

**Lemma 3 :**

$\forall i \geq 87, c(i) = c(i-34)$ *where $c(i)$ is the class of the row of length $i$.*

**proof :**

For $i \in [87, 176]$, we can verify by computing the row's classes that the lemma holds. Let $i \geq 176$ then at least one of the rows obtained by transition is greater then 86, then, by induction, we have an equivalence between playing on the row $i$ and on the row $i - 34$.

□

| Classes | Length | Length modulo 34 for rows of length > 53 |
|---|---|---|
| $C_0$ | 0,1,5,9,15,21,25,29,35,39,43, 55,59,63,73,77 | 5,9,21,25,29 |
| $C_1$ | 2,3,7,8,22,23,27,28,36,37,41, 42,56,57,61,62,70,71,75,76 | 2,3,7,8,22,23,27,28 |
| $C_2$ | 4,12,13,17,18,26,32,38,46,47, 52,60,72,80,81 | 4,12,13,26 |
| $C_3$ | 6,10,11,19,20,24,40,44,45,53, 54,58,66,74,78,79 | 6,10,11,19,20,24,32 |
| $C_4$ | 14,30,34,48,49,64,68,82,83 | 0,14,15,30 |
| $C_5$ | 16,31,50,51,65,84,85 | 16,17,31 |
| $C_6$ |  |  |
| $C_7$ | 33,67 | 33 |
| $C_8$ | 69 | 1 |
| $C_9$ | 86 | 18 |

Table 1: Row classes for the domino game

### 3.4.6 Classes of the white game:

Using the lemma 1 one can easily obtain the table of the row classes for the white game, by removing 1 from the domino-classes table.

## 4 The quantum game

If a player has a winning situation for the white game (considering only the white squares), he can force the other player to take the first grey square (by playing a winning strategy of the white game) and so to have at least 50%to win. If he doesn't play such a strategy, he can be forced to take the first grey and will have at most 50% to win. Therefore, a player can consider only the white squares, if he is in a winning situation for the white game, an optimal strategy consists in playing a white game winning strategy to force the other player to take the first gey square.

### 4.1 Alternate grey game

If we consider the particular case where we have only $n$ isolated gray squares, the probability of loosing of the first player evolve following the sequence $u_n$ where $n$ corresponds to the number of grey squares.

$$\begin{aligned} u_0 &= 1 \\ u_{n+1} &= 1/2 + 1/2(1 - u_n) \end{aligned}$$

The even subsequence is decreasing from 1 to 2/3.

$$\begin{aligned} e_0 &= u_0 = 1 \\ e_{n+1} &= u_{2(n+1)} = 1/2 + 1/4\ e_n \end{aligned}$$

The odd subsequence is increasing from 1/2 to 2/3.

$$\begin{aligned} o_0 &= u_1 = 1/2 \\ o_{n+1} &= u_{2(n+1)+1} = 1/2 + 1/4\ o_n \end{aligned}$$

We notice that if the number of grey squares is even(odd) then the probability that the first player looses is in $[2/3, 1]$ (in $[1/2, 2/3]$), thus we see that the parity of the number of gray is sufficient to have a framing of the loosing probability.

### 4.2 Lemma of the lucky player

Suppose that the players are lucky : at each time they choose a grey square it appears white so that the game stops only when there is no playable squares.

**Lemma 4 :**

 The parity of the number of grey squares played at the end of the lucky players' game is even for an odd white row and odd for an even white row.

**proof :**

 For a white row that has i white squares, let $Ng(i)$ denotes the set of possible number of grey squares played at the end of the game.

 When we play over a white row, we obtain rows that have a grey border square, therefore, in this proof, we call indifferently $n$ row : a row of $n$ white square that can have 0,1 or 2 grey border squares.

- For a 0 row (1 or two grey squares): $Ng(0) = 1$.

- For a 1 row (1 white square): We can have 0 or 2 grey squares played so : $Ng(1) = \{0, 2\}$ : contains only even numbers.

- For a 2 row (2 white squares): We can have 1 or 3 grey squares played so : $Ng(2) = \{1, 3\}$: contains only odd numbers.

- For an $l$ row, the possible squares taken are :

    A border grey square: we have one grey square played and a row of $(l-1)$ squares so $Ng(l)$ contains $1 + Ng(l-1)$, thus, by induction, the lemma holds.

    A border white square: $Ng(l)$ contains $Ng(l-2)$, thus the lemma holds.

A white square in the center of the row would create a $k$ row and a $l-k-3$ row: $Ng(l)$ contains $Ng(k)+Ng(l-k-3)$

If $l$ and $k$ have same parity $Ng(l-k-3)$ contains only even numbers.

Otherwise, it contains only odd numbers, thus the lemma holds.

Then $Ng(l)$ contains only odd numbers if i is even and even numbers if i is odd. □

## 4.3 Evolution of the general game:

We characterize a situation by a couple:

The first element is the parity of the number of grey squares played in a lucky players' game.

The second one is W for winning situations for the white game, L for loosing situations. The smallest situations for each couple are represented in figure 3

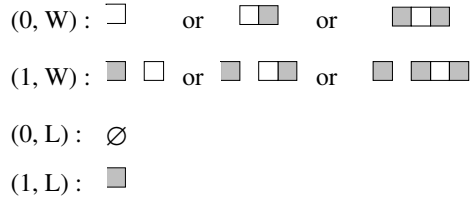

Figure 3: Smallest cases

We can represent in figure 4 the evolution of the game in the case that the players play well : a player that have a winning situation for the white game play a winning transition for this game, thus the new situation is loosing for the white game and have the same parity of grey squares taken in the lucky game. A red arrow corresponds to the choice of a grey square. We can notice that if a player takes a grey square the parity of the number of grey squares taken in the lucky game for the new situation change.

## 4.4 Loosing probability framing:

The loosing probabilities verify:

$$\begin{array}{rcccl} 0 & \leq & p(0,W) & < & 1/3 \\ 1/3 & < & p(1,W) & \leq & 1/2 \\ 2/3 & < & p(0,L) & \leq & 1 \\ 1/2 & \leq & p(1,L) & < & 2/3 \end{array} \qquad (2)$$

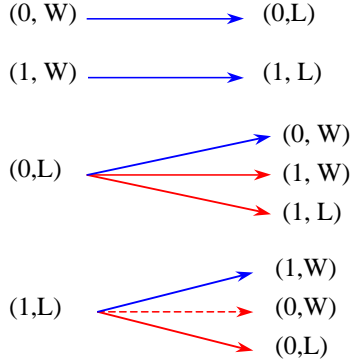

Figure 4: Evolution of the game

**proof :**

The framing holds for the smallest situations (figure 3). Suppose it's true for all situations that have number of playable squares less than $k$. Let $X$ be a situation for which this number is $k+1$,

If it is a $(0,W)$ situation: the loosing probability is $1-p(0,L) \in [0,1/3]$

If it is a $(1,W)$ situation: the loosing probability is $1-p(1,L) \in [1/3,1/2]$

If it is a $(0,L)$ situation: If the player choose a transition toward a $(0,W)$ situation: $p \in [2/3,1]$ If the player chooses a transition toward a $(1,W)$ situation: $p \in 1/2 + 1/2[1/2,2/3] = [3/4,5/6]$ If the player chooses a transition toward a $(1,L)$ situation: $p \in 1/2 + 1/2[1/3,1/2] = [2/3,3/4]$ So in all cases $p \in [2/3,1]$

If it is a $(1,L)$ situation: If the player chooses a transition toward a $(1,W)$ situation: $p \in [1/2,2/3]$ If the player chooses a transition toward a $(0,W)$ situation: $p \in 1/2 + 1/2[2/3,1] = [5/6,1]$ If the player chooses a transition toward a $(0,L)$ situation: $p \in 1/2 + 1/2[0,1/3] = [1/2,2/3]$ However, the transition toward a $(0,W)$ situation is never compulsory because a $(0,W)$ have at least one white square so the player could have chosen a $(1,W)$ transition (we can't create white squares). So $p \in [1/2,2/3]$. □

## 5 Generalization: Quantum octal games

### 5.1 Definition

The domino game belongs to the family of octal games that was introduced by Conway and Berlekamp in [2] : We have a set of rules that permit to take $i$ adjacent squares in the center of a row (creating two rows), in the border of a row (creating 1 row) and in a row of length $i$ (without creating any row). We can encode the allowed cases of taking $i$ adjacent squares with 3 bits: the $j^{th}$ bit will be 1 if the transition that generate $j$ rows is allowed.
Examples:

- 7=111 corresponds to the case that the three operations are allowed.

- 3=011 shows that we can take i adjacent cases only if we don't generate two rows.

So the rules can be written as an octal number where the $i^{th}$ number represents the rules corresponding to taking $i$ adjacent squares. For instance, we denote 0.07 the Domino game and 0.137 the White game.

More generally, we define Quantum octal games where we apply $\sqrt{Not}$ to adjacent squares of the choen block.

We denote $Q0.7$ the game studied in the previous section, where a player can take 1 qubit anywhere in a row.

A White game can be associated with any Quantum octal game, for instance, we associate the White game 0.137 with $Q0.7$, 0.0137 with $Q0.07$, 0.13 with $Q0.3$ and 0.03137 with $Q0.27$.

In deed, given a Quantum octal game $Q0.q_1q_2\ldots$, the associated White game $0.w_1w_2\ldots$ can be obtained as follows:

Let $b_j(a)$ the digit with weight $2^j$ in the binary decomposition of $a$.

- $b_0(w_i) = 1 \iff (b_0(q_i) = 1)\ or\ (b_1(q_{i-1}) = 1)\ or\ (b_2(q_{i-2}) = 1)$

- $b_1(w_i) = 1 \iff (b_1(q_{i-1}) = 1)\ or\ (b_2(q_{i-2}) = 1)$

- $b_2(w_i) = 1 \iff (b_2(q_{i-2}) = 1)$

**Lemma 5** :

If a quantum octal game is such that:

$$\begin{aligned} b_0(q_1) = b_1(q_1) = 1 \\ \forall i, q_{2i} = 0 \end{aligned} \quad (3)$$

Then the loosing probability framing (see section 4.4) still holds.

**proof** :

First, we prove that the lucky players' lemma holds: The parity of the number of grey squares played at the end of the lucky players' game is even for a row that contains an odd number of white squares and odd for a row that contains an even number of white squares.

The property is trivial for a block of 0 or 1 white square ($b_0(q_1) = b_1(q_1) = 1$).

Suppose that the player takes $p$ squares ( $p$ is odd) from a row that contains $l$ white squares:

- If the block chosen contains two grey squares, then $p = l + 2$ and $Ng(l) = 2$, thus the lemma holds ($l$ is odd).

- If the block chosen contains one grey square:

    If $p = l+1$, then $Ng(l) = 1$, thus the lemma holds ($l$ is even).

    Otherwise, $Ng(l)$ contains $1 + Ng(l-p)$ white squares, thus the lemma holds : $l - p$ has the opposite parity of $l$, and so by induction $Ng(l-p)$ has the same parity than $l$ and $Ng(l) = 1 + Ng(l-p)$ has an opposite parity than $l$.

- If the block chosen does not contain grey squares

    If $p = l$, then $Ng(l) = 0$ and $l$ odd.

    If it is a border block then $Ng(l)$ contains $Ng(l-p-1)$, thus the lemma holds.

    If it is a central block, it creates two rows, let $i$ denotes the number of white squares in the first row : $Ng(l)$ contains $Ng(i) + Ng(l-p-2-i)$

    If $l$ is odd, then $i$ and $l - p - i$ have the same parity and then, by induction, $Ng(l)$ is even.

    If $l$ is even, then $i$ and $l - p - i$ have the same parity and then, by induction, $Ng(l)$ is odd.

Thus the lemma of the lucky player still holds.

Now, we proof that the evolution of the game is analogous to the particular case studied in the previous section. As the lucky player lemma holds, we still can associate a couple to each situation of play, and the possible transitions are:

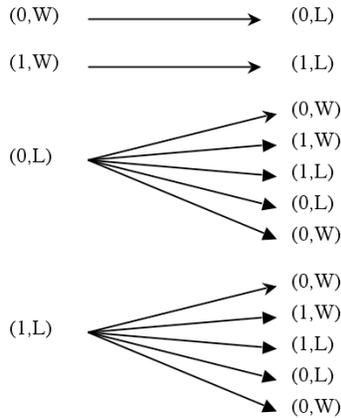

Figure 5: Evolution of the game

We note that we have new transitions that may appear, that are due to a possible choice of two grey squares. However, these transitions acts like the transition $(0, L) \longrightarrow (1, W)$ in the previous analysis: as $b_0(q_1) = b_1(q_1) = 1$, this choice is never compulsory. Moreover, it can not improve the winning probability of any situation so the loosing probability framing (see section 4.4) still holds. □